\documentclass[10pt,final,journal]{IEEEtran}
\usepackage{cite}
\usepackage{url}
\usepackage[dvipsnames]{xcolor}
\usepackage{bbding}
\usepackage{hyperref}

\usepackage{multirow} 
\usepackage{amsmath}
\usepackage{booktabs}
\usepackage{makecell}
\usepackage{amsmath}
\usepackage{amssymb}
\usepackage{times}
\usepackage{epsfig}
\usepackage{graphicx}
\usepackage{float}

\usepackage{pifont}
\usepackage{color}
\usepackage{cite}
\usepackage{xcolor,graphicx,float}
\graphicspath{{./figures/}}
\usepackage[numbers,sort&compress]{natbib}
\usepackage{amsmath}
\usepackage{bm}
\usepackage{soul}
\newcommand{\etal}{\textit{et al}.}
\newcommand{\ie}{\textit{i}.\textit{e}.}
\newcommand{\eg}{\textit{e}.\textit{g}.}
\usepackage{ragged2e}

\ifCLASSOPTIONcompsoc
\usepackage[caption=false,font=normalsize,labelfont=sf,textfont=sf]{subfig}
\else
\usepackage[caption=false,font=footnotesize]{subfig}
\fi

\begin{document}

%

\title{Perception-Distortion Balanced Super-Resolution: \\ A Multi-Objective Optimization Perspective}
%
%

\author{Lingchen~Sun,
        Jie~Liang, Shuaizheng~Liu,
        Hongwei~Yong, Lei~Zhang,~\IEEEmembership{Fellow, IEEE} 
\thanks{L.~Sun, S.~Liu, H.~Yong, and L.~Zhang are with the Department of Computing, the Hong Kong Polytechnic University, Hong Kong (e-mail: ling-chen.sun@connect.polyu.hk;
shuaizhengliu21@gmail.com;
hongwei.yong@polyu.edu.hk;
cslzhang@comp.polyu.edu.hk). This work is supported by the Hong Kong RGC RIF grant (R5001-18) and the PolyU-OPPO Joint Innovation Lab.}
\thanks{J.~Liang is with the OPPO Research Institute (e-mail: liang27jie@gmail.com).}}

\maketitle

\begin{abstract}
High perceptual quality and low distortion degree are two important goals in image restoration tasks such as super-resolution (SR). Most of the existing SR methods aim to achieve these goals by minimizing the corresponding yet conflicting losses, such as the $\ell_1$ loss and the adversarial loss. Unfortunately, the commonly used gradient-based optimizers, such as Adam, are hard to balance these objectives due to the opposite gradient decent directions of the contradictory losses. In this paper, we formulate the perception-distortion trade-off in SR as a multi-objective optimization problem and develop a new optimizer by integrating the gradient-free evolutionary algorithm (EA) with gradient-based Adam, where EA and Adam focus on the divergence and convergence of the optimization directions respectively. As a result, a population of optimal models with different perception-distortion preferences is obtained. We then design a fusion network to merge these models into a single stronger one for an effective perception-distortion trade-off. Experiments demonstrate that with the same backbone network, the perception-distortion balanced SR model trained by our method can achieve better perceptual quality than its competitors while attaining better reconstruction fidelity. Codes and models can be found at \href{https://github.com/csslc/EA-Adam}{https://github.com/csslc/EA-Adam}.
\end{abstract}


%
\IEEEpeerreviewmaketitle

\section{Introduction}
%
%
%
%
\IEEEPARstart{I}mage super-resolution (SR) \cite{9044873} is a valuable technique to reconstruct a high-resolution (HR) image with good quality from a low-resolution (LR) input image. Benefiting from the deep learning techniques, it has become popular to train a deep neural network (DNN) for SR by using a large amount of LR-HR image pairs \cite{srcnn,fastsrcnn,sajjadi2017enhancenet, haris2018deep, kim2016deeply,srgan,esrgan, laroche2023deep}. 
One commonly used loss function to optimize the SR models is the distortion (or fidelity) measure, \eg, the norm of the prediction errors \cite{srcnn, zhang2018image, anwar2020densely, denseSR}. However, SR is a typical ill-posed problem, and there are many possible HR estimations for the given LR input. It is well-known that the $\ell_2$ or $\ell_1$ loss tends to generate over-smoothed HR estimates \cite{LDL, hinders} although they can result in good distortion measures such as peak-to-noise ratio (PSNR). 

To better preserve the local structures and details of reconstructed images, the SSIM loss \cite{ssim} and perceptual loss \cite{perceptualloss} have also been used to train SR models. The former adopts the widely used image quality assessment index, \ie, SSIM \cite{ssim,ms-ssim}, as the optimization objective, while the latter minimizes the $\ell_1$ or $\ell_2$ norm of the prediction difference in a feature space \cite{srgan,esrgan}. In general, the SSIM loss and perceptual loss can improve the perceptual quality of SR outputs, but the improvement is limited because they can hardly generate details that are lost during degradation\cite{pdtradeoff,soh2019natural}.

\begin{figure}
	\centering 
	\includegraphics[scale=0.9]{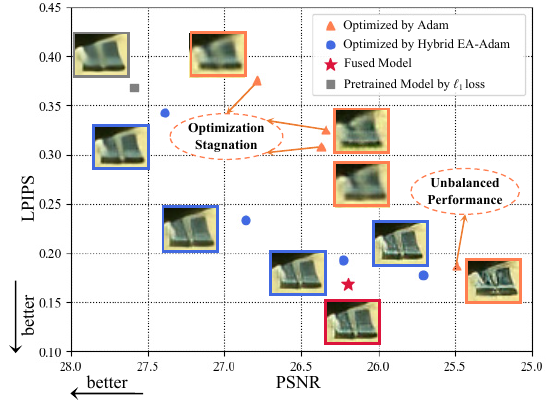}
	\caption{Performance comparison of perception-distortion balanced SR models trained by the Adam optimizer and our hybrid EA-Adam optimizer on the BSD100 dataset. For the perceptual quality index LPIPS, the smaller the better. For the distortion degree index PSNR, the higher the better. The models optimized by Adam (orange triangle points) with different weighted combinations of $\ell_1$, perceptual and adversarial losses are either stagnated into an unsatisfactory solution, or achieve unbalanced performance. The models optimized by our EA-Adam optimizer (blue circle points) can obtain better convergence and divergence. Furthermore, the fused model (red star) from EA-Adam models achieves the best perceptual quality index while maintaining a high fidelity index.}		
	\label{fig1}

\end{figure}

To obtain perceptually more realistic HR images, the perception-distortion trade-off methods \cite{2018ECCVW,srgan,esrgan,usrgan,ranksrgan, 9727093} have been developed in SR.
Usually, they train a generative adversarial network (GAN) \cite{gan} by using the adversarial loss. Instead of minimizing the pixel-wise $\ell_2$ or $\ell_1$ loss or local structure-wise SSIM or perceptual loss, the adversarial loss is optimized to minimize the distribution divergence between the predicted SR images and HR images. In practice, it is unstable and impossible to train the SR model using only the adversarial loss. Therefore, most GAN-based SR methods weigh the $\ell_1$, perceptual and adversarial losses \cite{srgan,esrgan} into one single loss for optimization.    

While GAN-based SR methods can generate some realistic details, they will also introduce undesired artifacts \cite{prashnani2018pieapp, pdtradeoff}. Many following works \cite{fuoli2021fourier, LDL, bestbeddy,zhang2022perception, DeSRA, SROOE, DualFormer} aim to reduce the artifacts caused by the adversarial loss while keeping the realistic details. Dario \etal ~\cite{fuoli2021fourier} applied $\ell_1$ and adversarial losses in both spatial and Fourier domains to improve the restoration of high-frequency details. Liang \etal ~\cite{LDL} designed an LDL loss to suppress GAN-generated artifacts by considering their statistical difference from visually friendly details. To avoid artifacts in low-frequency areas, a region-aware adversarial loss \cite{bestbeddy} was designed to help the discriminator pay more attention to high-frequency areas. Zhang \etal ~\cite{zhang2022perception} used a similar idea to handle the low-frequency and high-frequency areas differently by using a constrained loss. Xie \etal ~\cite{DeSRA} proposed to detect the
artifact regions and developed a finetuning procedure to improve GAN-based SR models.

Unfortunately, the above-mentioned methods have two major limitations. First, empirically selecting weights to combine the losses into one is unreasonable and labor-consuming \cite{zhang2022perception}. Second and more importantly, distortion- and perception-oriented losses often conflict with each other, and the commonly used gradient-based DNN optimizers, such as Adam \cite{adam}, are difficult to balance the conflicting gradient-decent directions. We illustrate this issue in Fig. \ref{fig1}, where we train a population of SR networks by weighing the $\ell_1$, perceptual and adversarial losses with different weights. The Adam optimizer is used to train the network, and the PSNR (the higher the better) and LPIPS (the smaller the better) \cite{lpips} indices are used to measure the distortion degree and perceptual quality, respectively. An ideal SR model is expected to achieve both high PSNR and low LPIPS, \ie, approaching the origin point in Fig. \ref{fig1}. However, we can see that the learned models (orange triangle points) by Adam with different weighted $\ell_1$, perceptual and adversarial losses are rather stagnated into an unsatisfactory solution, or achieve unbalanced performance. 

In this paper, we propose a new approach for perception-distortion balanced SR from the network optimization perspective. We formulate the problem as a multi-objective optimization task \cite{pdtradeoff}, one objective aiming for high perceptual quality and another for low distortion degree. Considering that these two objectives are hard to achieve simultaneously by using gradient-based optimizers because of their conflicting gradient descent directions, we propose to integrate the gradient-free evolutionary algorithm (EA) \cite{nsga,nsgaII,moead} with the gradient-based Adam for network optimization. 
Adam mainly ensures the model convergence, \ie, the losses will decrease along the gradient, while EA mainly ensures the model divergence, \ie, the potential solutions could interact with and benefit from each other by the crossover operation. Meanwhile, the mutation operation of EA helps the model to jump out of the local optimum. 
As a result, the obtained Pareto front (PF) consists of a series of optimal models with different perception-distortion preferences, as shown by the blue circular points in Fig. \ref{fig1}. Compared with the Adam optimizer, the models optimized by our proposed EA-Adam method are closer to the original point and are more evenly distributed than those optimized by Adam.
Motivated by the 
mixture of experts theory~\cite{moe, DASR} and the network interpolation methods \cite{network-interpolation, liu2022adaptive},
we further propose a simple yet effective fusion network to merge the obtained models into a single stronger one, which inherits the advantages of the model population.
As indicated by the red star in Fig. \ref{fig1}, the fused model achieves the best perceptual quality index while maintaining a high fidelity index. 
Extensive experiments on benchmark datasets demonstrate the effectiveness of the proposed hybrid EA-Adam optimizer and weight aggregation network. Compared with the existing state-of-the-art methods, our model achieves a better perception-distortion trade-off.

The rest of this paper is organized as follows. Section II summarizes the related works on distortion- and perception-oriented SR methods, and evolutionary algorithms on DNN learning. Section III presents the multi-objective formulation of SR, the hybrid EA-Adam optimizer, and the network fusion approach in detail. Section IV reports the experimental results and discussions. Finally, Section V concludes this paper.


\section{Related Work}
\subsection{Distortion-oriented SR Methods}
Most of the SR methods in this category \cite{srcnn, fastsrcnn,anwar2020densely,haris2018deep} employ the pixel-wise loss functions, such as the $\ell_2$ or $\ell_1$ norm of the prediction error, while focusing on the design of network architectures to improve the SR performance. The seminal work of SRCNN \cite{srcnn} introduced a convolutional neural network (CNN) with three layers to perform the SR task. Afterward, many SR models have been developed. Kim~\etal~\cite{kim2016deeply} trained a deeper and wider network to improve the representational ability of CNN. 
Deep residual connections were introduced to improve deep SR network learning performance \cite{lim2017enhanced, zhang2018residual,zhang2018image}. Tong~\etal~\cite{denseSR} proposed to use dense skip connections in a very deep network.
The recursive network~\etal~\cite{kim2016deeply} was designed to improve SR performance without introducing new parameters for additional convolutions.
Very recently, the transformer networks \cite{swinir,elan,hatSR, 10140179} have been successfully employed for SR, resulting in better distortion-based measures such as PSNR.
Because $\ell_2$ or $\ell_1$ loss tends to generate blurry SR outputs, the SSIM loss \cite{ssim} and perceptual loss \cite{perceptualloss} have been proposed to regularize the local structures of images. These two losses can improve the visual quality of SR outputs without harming too much the distortion measures. However, they cannot generate image details to make the SR image perceptually more realistic. 

\subsection{Perception-oriented SR Methods}
To make the reconstructed SR images perceptually more realistic, SRGAN \cite{srgan} combined the adversarial loss with distortion-oriented losses (\eg, $\ell_1$ loss) in training. Following SRGAN, ESRGAN \cite{esrgan} used the VGG features before activation and employed the RRDB backbone \cite{esrgan} as the generator model, improving the perceptual quality. DualFormer \cite{DualFormer} leveraged the spectral and spatial discriminators for identifying high-frequency and low-frequency information, respectively. While GAN-based methods can generate new image details, they may introduce many unnatural visual artifacts because of unstable adversarial training. 
Therefore, many of the following works aim to reduce the GAN-generated artifacts from the perspectives of frequency division \cite{bestbeddy,fuoli2021fourier,zhang2022perception,huang2017wavelet, WGSR}, the weight of layers in perceptual loss \cite{SROOE, park2022flexible}, or different statistical properties \cite{LDL, DeSRA}. In addition, some works \cite{ranksrgan} aim to optimize the no-reference image quality metrics, which can be non-differentiable. For example, RankSRGAN \cite{ranksrgan} designed a differentiable rank-content loss to rank the quality of predictions with NIQE \cite{niqe} or Ma \cite{ma} index.

\subsection{Evolutionary Algorithms in DNN Learning}
EA \cite{nsgaII,moead} is a classical search-based optimization technique, where individuals interactively evolve from a population. The EA operators mainly consist of crossover \cite{sbx}, mutation, and selection. Due to the global search and gradient-free properties, EA has proven its effectiveness in many applications \cite{zhou2021survey, saravanan2009evolutionary, mala2018lost}. It has recently been applied to DNN learning \cite{guo2020single,yu2020bignas,eagan}. Guo~\etal~\cite{guo2020single} applied evolutionary architecture search to construct a simplified supernet with all architectures optimized simultaneously. Yu~\etal~\cite{yu2020bignas} searched  over a big single-stage model that contains small children models. EAGAN \cite{eagan} applied an efficient two-stage EA-based network search framework to learn GANs. Another application is network compression. For example, Zhou~\etal~\cite{zhou2019knee} established filter pruning as a multi-objective optimization problem and proposed a knee-guided EA algorithm to trade-off between the scale of parameters and performance. In addition, EA algorithm can be used as the network optimizer. ESGD \cite{evolutionaryopt} alternately applied SGD and EA in one framework to improve the average performance of DNN. Zhang~\etal~\cite{zhang2022optimizing} proposed a hierarchical cluster-based suppression algorithm to remove similar weights and improve population diversity. GEMONN \cite{gradienteaopt} determined the search direction in the EA optimizer by the gradient of weights to improve the efficiency of EA for training DNN.

To optimize a network, EA is usually coupled with a gradient-based optimizer to solve large-scale optimization problems \cite{tian2022integrating}. Most of the previous EA-based DNN learning works \cite{evolutionaryopt} optimize a single objective. In this work, we formulate the perceptually realistic SR task as a multi-objective optimization problem and propose a hybrid EA and Adam algorithm to solve it.

\section{The Proposed Method}


\begin{figure*}
	\centering 
	\includegraphics[scale=0.61]{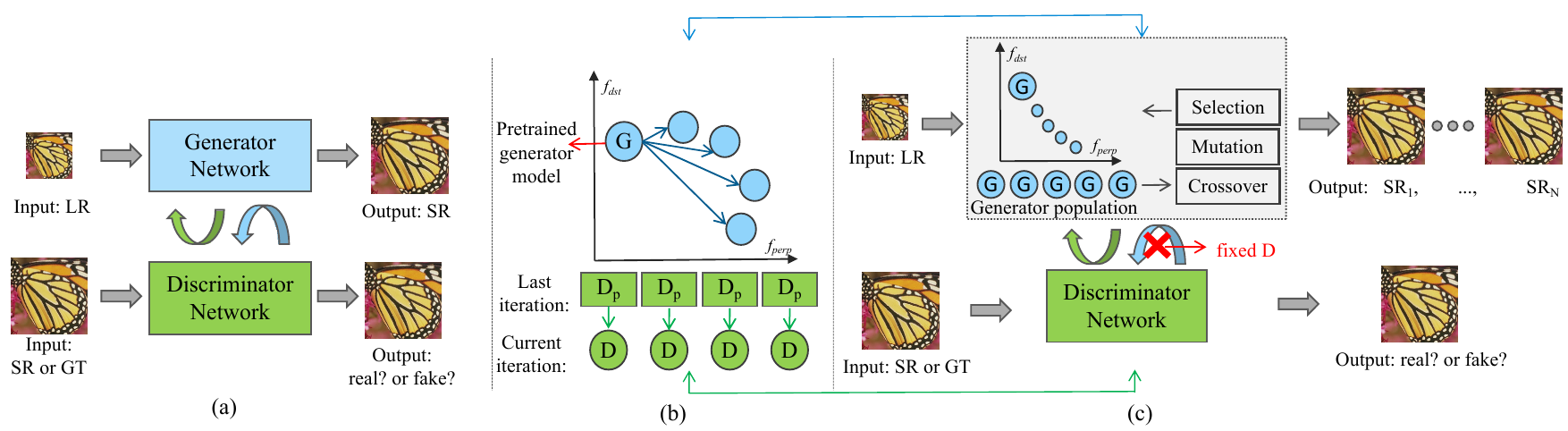}
	\caption{(a) The framework of traditional GAN-based SR methods. (b) The Adam steps optimize a set of generator and discriminator networks. (c) The EA steps optimize the generator networks by fixing the discriminator networks.}	
	\label{fig3}
\end{figure*}

\label{sec:intro}
This section presents the proposed hybrid EA-Adam optimizer and a simple yet effective fusion network for perception-distortion balanced SR. Firstly, we formulate the SR task as a multi-objective optimization problem, with two objectives focusing on perceptual quality and fidelity quality, respectively. Secondly, a hybrid EA-Adam optimizer is designed to solve the problem, where EA addresses the model divergence and Adam addresses the model convergence. A set of SR models with different perception-distortion preferences are trained to form a PF. Lastly, a simple yet effective fusion network is proposed to merge the learned models into a strong perception-distortion-balanced SR model.

\subsection{Multi-Objective Formulation}

Most of the GAN-based SR methods \cite{srgan,esrgan} train the generator model by weighting the pixel-wise $L_{pix}$ loss, perceptual loss $L_{percep}$ and adversarial loss $L_{adv}$ as follows:
\begin{equation}
    {L_{GAN}} = {\alpha_1}{L_{pix}} + {\alpha_2}{L_{percep}} + {\alpha_3}{L_{adv}},
    \label{Eq1}
\end{equation}
where the $L_{pix}$ (\ie, $\ell_1$ loss) optimizes the reconstruction fidelity, and $L_{percep}$ and $L_{adv}$ improve the perceptual quality. The weights $\alpha_1$, $\alpha_2$ and $\alpha_3$ are often empirically set to 0.01, 1 and 0.005, respectively. 

The Adam optimizer \cite{adam} is dominantly used to optimize the above loss function. Unfortunately, the gradient-based Adam is difficult to minimize the perception- and distortion-oriented losses simultaneously due to their opposite gradient descent directions, while an ideal weight setting is hard to be found given an infinite searching space. Therefore, most traditional SR models turn to pursue either fidelity or perceptual quality yet sacrifice the other. As illustrated in Fig. \ref{fig1}, when the weights are biased towards distortion- or perception-oriented losses, the SR results will be over-smoothed or include visually unpleasant artifacts. When similar weights are set to balance perception and distortion, the optimization might be stagnant at the local minimum as no explicit gradient descent directions can be found. Note that we apply gradient normalization \cite{chen2018gradnorm} to avoid the trivial solution.

In this paper, rather than linearly combining the distortion- and perception-oriented losses into a single objective function, we formulate the perceptually realistic SR task as a multi-objective optimization problem as follows:
\begin{equation}
\left\{ \begin{array}{l}
\min {f_{1}} = {L_{pix}},\\
\min {f_{2}} = {L_{percep}} + {\alpha}{L_{adv}},
\end{array} \right.\label{Eq2}
\end{equation}
where objectives $f_1$ and $f_2$ respectively focus on the reconstruction fidelity and the perceptual quality of the SR results. In the following, we propose to couple EA \cite{moead} with Adam \cite{adam} to address the multi-objective optimization problem in Eq.~\eqref{Eq2}. EA and Adam help with the optimization process for the divergence and convergence, respectively.

\begin{table}

\caption{
\begin{justify}
\justifying
The proposed hybrid EA-Adam optimization algorithm.
\end{justify}
}
\vspace{3pt}
 \label{table1}
	\centering
	\begin{tabular}{r}
		\hline
		\hline
		\specialrule{0em}{1pt}{1pt}
		\small
		\parbox[c]{8cm}{ \textbf{Input}: The number of optimized models $N$, Adam epochs $T^{Adam}$, EA epochs $T^{EA}$, total epochs $T$, the pretrained model ${\theta}_{G_0}$ optimized by $\ell_1$ loss, and the probability $\delta$ of selecting parents from the population $\mathcal{B}$.} \\
		\quad\\
		\small
        \parbox[c]{8cm}{\textbf{Step 1. Initialization}: Initialize a population of generators $\mathcal{G} = \{\theta_{G_1}, \cdots, \theta_{G_N}\}$, where ${\theta}_{G_k} =  {\theta}_{G_0}, k\in [1, \cdots, N]$. Initialize the discriminator population ${\mathcal{D}}= \{\theta_{D_1}, \cdots, \theta_{D_N}\}$. Generate the neighboring population $\mathcal{B}$. Define the Adam optimizers for each model.} \\
		\quad\\
		\small
		\parbox[c]{8cm}{\textbf{Step 2. Optimization iteration}:} \\
		\small
		\parbox[c]{8cm}{\textbf{For} $t \in 1:T^{EA}+T^{Adam}:T$ \textbf{do}}\\
             \small
		\parbox[c]{7.5cm}{$\#$Adam Steps}\\
		\small
		\parbox[c]{7.5cm}{\textbf{For} $i \in \{1,\cdots,T^{Adam}\}$ \textbf{do}}\\
		\small
		\parbox[c]{7cm}{\textbf{For} $s \in \{2,\cdots,N\}$ \textbf{do}}\\
		\small
		\parbox[c]{6.5cm}{\textbf{1.} Update the weights of $\theta_{G_s}$ by Adam.}\\
		\small
		\parbox[c]{6.5cm}{\textbf{2.} Update the weights of $\theta_{D_s}$ by Adam.}\\
		\small
		\parbox[c]{7cm}{\textbf{EndFor}}\\
		\small
		\parbox[c]{7.5cm}{\textbf{EndFor}}\\
		\small
		\parbox[c]{7.5cm}{$\#$EA Steps}\\
		\small
		\parbox[c]{7.5cm}{\textbf{For} $j \in \{1,\cdots,T^{EA}\}$ \textbf{do}}\\
		\small
		\parbox[c]{7cm}{\textbf{For} $k \in \{2,\cdots,N\}$ \textbf{do}}\\
		\small
		\parbox[c]{6.5cm}{\textbf{1}. If $r<\delta$,  $\mathcal{P}=\mathcal{B}(k)$; otherwise, $\mathcal{P}=\mathcal{G}$.}\\
		\small
		\parbox[c]{6.5cm}{\textbf{2}. Select parents ${\theta}_1$ and ${\theta}_2$ from $\mathcal{P}$  randomly.}\\
		\small
		\parbox[c]{6.5cm}{\textbf{3}. Generate offspring ${\theta}_I$ by the crossover and mutation in Eq.~\eqref{Eq3} and Eq.~\eqref{Eq4}.}\\
		\small
		\parbox[c]{6.5cm}{\textbf{4}. Select the new individual $\theta_{G_k}$ as in Eq.~\eqref{Eq5}.}\\
		\small
		\parbox[c]{7cm}{\textbf{EndFor}}\\
		\small
		\parbox[c]{7.5cm}{\textbf{EndFor}}\\
		\small
		\parbox[c]{7.5cm}{\textbf{Re-initialization:} Re-define the Adam optimizer for each model in population.}\\
		\small
		\parbox[c]{8cm}{\textbf{EndFor}}\\
		\hline
		\hline
	\end{tabular}
\vspace{-3pt}
\end{table}

\begin{figure}[t]
	\centering 
	\includegraphics[scale=0.5]{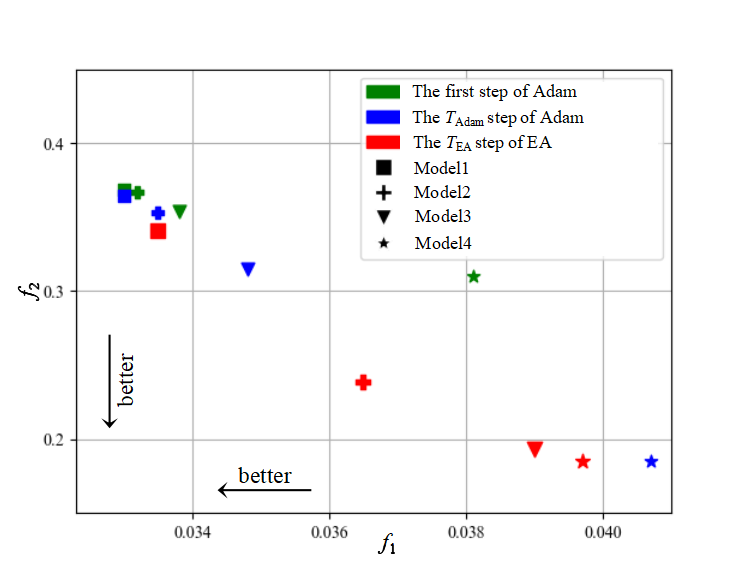}
	\caption{Illustration of the convergence process of the proposed EA-Adam optimizer in an EA-Adam cycle ($T_{Adam}+T_{EA}$). The Adam steps primarily focus on optimization convergence, enabling rapid search for improved objective losses. In contrast, the EA step emphasizes divergence optimization, ensuring a uniform spread of solutions across the plane.}
	\label{convergence}
\end{figure}

\begin{figure*}[t]
	\centering 
	\includegraphics[scale=0.55]{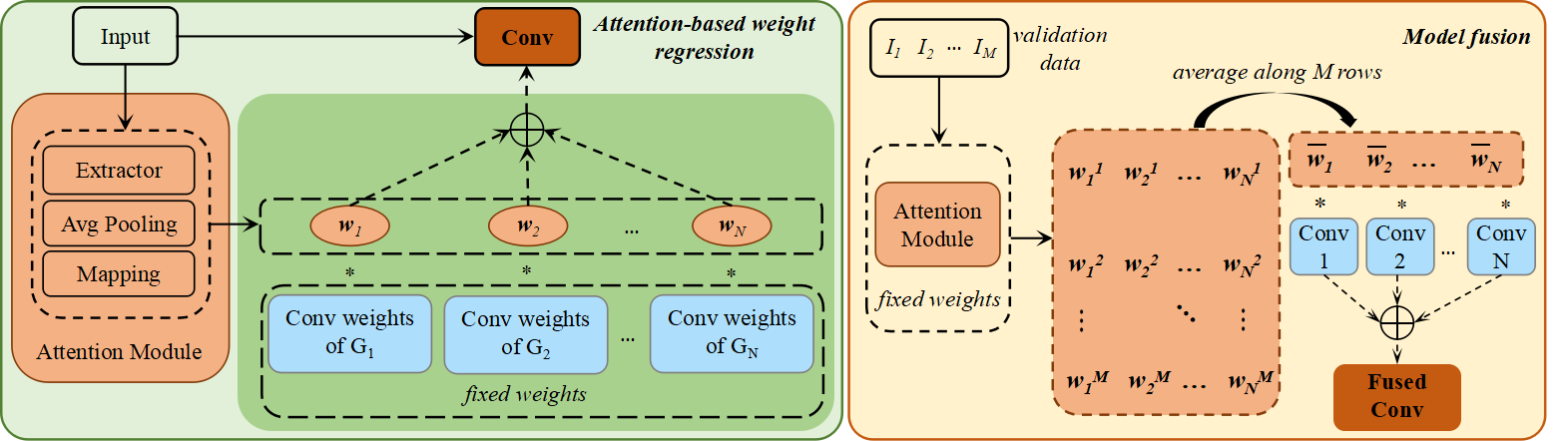}
	\caption{Two stages of our network fusion process. Left: training of the attention-based weight regression network, which predicts the fusion weights of experts for each input image. Right: model fusion by averaging weight vectors of a batch of validation data.}	        
	\label{fig4}
\end{figure*}

\subsection{The Hybrid EA-Adam Optimizer}

The algorithm of the proposed hybrid EA-Adam optimizer is summarized in Table \ref{table1}. EA and Adam alternately optimize the $N$ SR models with different perception-distortion preferences. We employ the general framework of GAN-based SR learning, which is shown in Fig. \ref{fig3}(a). The generator network learns the mapping between the LR and HR images, and the discriminator network discriminates between the true and fake images. The two networks compete with each other to ensure that the generated images are of high quality and more natural.

The Adam step is illustrated in Fig. \ref{fig3}(b). Initialized by a pre-trained model with $\ell_1$ loss, the other $N-1$ generator-discriminator pairs are updated by Adam separately, as shown in the Adam step of Table \ref{table1}. We fix the first model across the training process as it has the best reconstruction fidelity by training with pure pixel-wise $\ell_1$ loss. Then, the EA step starts, as illustrated in Fig. \ref{fig3}(c). The adversarial training facilitates the model convergence in the Adam step to obtain better objective scores, while the EA step focuses on the model divergence, enabling the optimized models uniformly distributed on the performance plane spanned by the two objective indices. Therefore, the discriminators are only updated in the Adam steps to enhance convergence.

In the EA step, a population of generator models is optimized together. Following MOEA-D \cite{moead}, the parents are first chosen from the neighboring population $\mathcal{B}$ or the whole population $\mathcal{G}$, which focuses on the local or global evolution. The offspring is generated by the selected parents via crossover and mutation operators. The crossover is used to interact with the individuals in a population with each other so that useful information can be effectively propagated internally. We employ the SBX crossover \cite{sbx}, which is defined as follows:
\begin{equation}
{\theta}_I = 0.5 \times \left[ {(1 + \beta ){\theta}_1} + (1 - \beta ){{\theta}_2} \right].
\label{Eq3}
\end{equation}
Here, $\{\theta\}_{I, 1, 2}$ indicate the parameters of the corresponding models, $\beta$ is a weighting variable that relies on a random number $r\in[0,1]$, where $\beta  = {{{(r \times 2)}^{1/(1 + \eta )}}}$ if $r<0.5$ and $\beta  = {{{(1/(2 - r \times 2))}^{1/(1 + \eta )}}}$ otherwise, $\eta$ is a constant and is set to 20 as in SBX~\cite{sbx}. The mutation operator disturbs the optimization to jump out of the local optimum by adding a zero-mean random Gaussian noise ${\varepsilon ^{(k)}}\sim {\mathcal{N}(0,0.01)}$ \cite{evolutionaryopt} to each of the $N$ models, \ie,
\begin{equation}
{\theta_I} = {\theta_I} + {\varepsilon ^{(k)}}, k\in\{1, \cdots, N\}.
\label{Eq4}
\end{equation}

To select a good model from the parent models and the newly generated offspring models, we measure the model performance by using an aggregation function derived from the Tchebycheff decomposition \cite{miettinen2012nonlinear} as follows:
\begin{equation}
\min {F^{te}} = \max \{ \lambda ({f_1} - z_1^*),(1 - \lambda )({f_2} - z_2^*)\}, \label{Eq5}
\end{equation}
where $z_1^*$ and $z_2^*$ are the moving minimum values of $f_1$ and $f_2$ recorded until the current iteration, and weights $\lambda$ are uniformly distributed within $(0,1]$. Specifically, we set $\lambda =  {0,\frac{1}{{N-1}}}, {\frac{2}{{N-1}},\cdots, 1}$ for the $N$ models, respectively. With different aggregation loss weights $\lambda$, the multiple optimization problem can be divided into $N$ sub-problems of single-objective optimization for different perception-distortion preferences. Note that the momentum in the Adam optimizer should be initialized after the EA optimization step. Finally, a set of generator models with different perception-distortion preferences are obtained to form a PF, as shown in Fig. \ref{fig1}.

To illustrate the training convergence
process of the EA-Adam optimization algorithm, in Fig. \ref{convergence} we demonstrate the performance of optimized models on the two conflicting objectives $f_1$ and $f_2$ over an Adam-EA cycle on the B100 dataset. Different shaped symbols represent the $N=4$ optimized models, where the green color denotes the model optimized after the first step of Adam, the blue color denotes the model optimized after $T_{Adam}$ steps by Adam ($T_{Adam} = 10$ in this experiment), and the red color denotes the model further optimized by EA with $T_{EA} = 1$ step. 
We can see that Adam focuses on the convergence of the 4 models, enabling a fast search for better objective scores of them. However, due to the opposite optimization gradient directions of $f_1$ and $f_2$, Adam suffers from optimization stagnation. That is, the optimized models targeting at different perception-distortion objectives tend to converge on the same side of the plane, making it difficult to form a diverse solution space. Fortunately, the introduction of EA enhances the diversification of the $N$ models so that the optimized models are more evenly distributed in the optimization plane with improved performance.



\subsection{Network Fusion}

As mentioned above, the $N$ SR models (with the same architectural topology) optimized by the proposed hybrid EA-Adam optimizer can work better than their counterparts optimized by Adam in balancing perceptual quality and distortion degree. 
However, it can be labor-consuming to manually select one from them to super-resolve the input LR image. On the other hand, if we apply all the $N$ models to the input image and then fuse the outputs into one image, it can be expensive and less effective in inference \cite{network-interpolation}. 
To solve this issue, we propose to automatically fuse the $N$ models into one stronger model to pursue further enhanced performance beneath the PF formed by the $N$ models. To achieve this goal, we first introduce a lightweight attention-based weight regression network to fuse the convolution layers of the $N$ SR models as in~\cite{DASR, moe, dynamic-conv}. As shown at the left of Fig. \ref{fig4}, the input is the feature from one layer and the output is the weight of the corresponding layer in the $N$ models. The generated weights satisfy $\sum\nolimits_k {{w_k} = 1} ,k\in\{1, \cdots, N\}$. There are 3 modules in the weight regression network. The first module consists of 3 convolution layers with Leaky ReLU activation. The kernel sizes of convolution layers are all $5 \times 5$. The second module is a global average pooling layer, and the last is a mapping module, which consists of two linear layers followed by a Sigmoid activation. The first module extracts the features and the last two modules estimate the weights to fuse the convolution layers of the SR models.

We train the attention-based weight regression network with the most commonly-used $L_{GAN}$ loss as it encourages the output to be perceptually more realistic. Meanwhile, since the aggregated model linearly interpolates among the $N$ SR models that are fixed in this stage, the reconstruction fidelity can also be well preserved.
We fuse the $N$ SR models in one shot and apply the fused model to all testing images, as shown at the right of Fig. \ref{fig4}. Here, we employ a set of $M$ validation images, \ie, $\{I_i\}_{i=1}^M$, to facilitate the fusion process. Based on the validation data, we calculate the adaptive fusion weights for each $I_i$, \ie, $[w_1^i, w_2^i, \cdots, w_N^i]$, using the trained attention-based weight regression network, and average them over the $M$ weight vectors to get a universal weight, \ie, $[\bar{w}_1, \bar{w}_2, \cdots, \bar{w}_N]$. Finally, a fused model is obtained, which inherits the advantages from the $N$ SR models in terms of both reconstruction fidelity and perceptual quality, while holding a stronger generalization capability. 

\section{EXPERIMENT}

\subsection{Experiment Setup}
\noindent\textbf{Datasets and evaluation metrics}. To train the population of SR models, we use either the DIV2K dataset \cite{div2k} (800 images) or the DF2K dataset \cite{div2k, flick2k} (800 DIV2K images and 2650 Flickr2k images)  according to the training datasets of competing methods. For the network fusion process, 700 images of DIV2K are used to train the weight regression network, and the remaining 100 images are used as the validation data in the network fusion process. Note that we do not employ any test data in the stages of network fusion process. Once learned, our weight regression network is fixed and applied to any given test set.

Among the 800 images in the DIV2K training dataset, we utilize 700 images as training data for the weight regression network, and the remaining 100 images are used as the validation data in the network fusion process. We have not employed any test data during the training stage. Once learned, our weight regression network is fixed and applied to any given test set. Therefore, our method is zero-shot, which aligns with the approach taken by the referenced and compared SR methods.  

Following prior arts \cite{esrgan,LDL,WGSR}, we conduct experiments with a scaling factor
of $4\times$ on both synthetic (downsampled using MATLAB bicubic kernel) and real-world (degraded using RealESRGAN pipeline \cite{realesrgan}) experiments.
For the bicubic-degraded SR task, we evaluate the performance of different methods on 6 benchmarks, including Set5 \cite{set5}, Set14 \cite{set14}, BSD100 \cite{b100}, Urban100 \cite{urban100}, Manga109 \cite{manga}, and DIV2K100 \cite{div2k}. We compute the PSNR and SSIM \cite{ssim} indices on the Y channel in the YCbCr space to measure the distortion degree, and compute the LPIPS \cite{lpips} and DISTS \cite{dists} indices in the RGB space to evaluate the perceptual quality. For the real-world-degraded SR task, we evaluate the performance of different methods on the RealSR \cite{realsr} and DrealSR \cite{drealsr} benchmarks.

\textbf{Backbones and compared methods:} As in many previous perceptually realistic SR works \cite{ranksrgan,usrgan,LDL,bestbeddy}, we adopt SRResNet \cite{srgan} and RRDB \cite{esrgan} as the backbone networks. SRResNet is a lightweight network proposed in SRGAN \cite{srgan}, while RRDB \cite{esrgan} is widely used in later GAN-based SR methods \cite{usrgan,spsr, LDL} for its superior performance.
Specifically, we compare our SRResNet-based model with SRGAN \cite{srgan} and RankSRGAN \cite{ranksrgan}, and compare our RRDB-based model with ESRGAN \cite{esrgan}, SPSR \cite{spsr}, LDL \cite{LDL}, CAL-GAN \cite{iccv2023} and WGSR \cite{WGSR}. 
In addition, Transformer-based backbones have become popular as they can capture long range dependency. Therefore. we compare SwinIR \cite{swinir} trained with the $L_{GAN}$ loss by Adam optimizer and our proposed hybrid EA-Adam optimization method to validate the effectiveness of our method on the Transformer backbone.
For the methods that apply different backbones or training datasets, we compare them with our model of similar capacity and training datasets. Specifically, we compare our SRResNet-based model with CFSNet \cite{cfsnet} and G-MGBP \cite{2018ECCVW}, and compare our RRDB-based model with SROOE\cite{EPSR}, PD-ADMM \cite{zhang2022perception} and DualFormer \cite{DualFormer}. 
We further validate the proposed method for the real-world SR tasks, and compare the obtained `RealESRGAN+Ours' model with RealESRGAN \cite{realesrgan}, IKC \cite{IKC}, IKR-Net \cite{IKR-Net}, LDL \cite{LDL} and DASR \cite{DASR}.
The results of the compared methods are obtained using the officially released models.
Note that we do not compare our method with the distortion-oriented SR methods, because this work focuses on perception-distortion balance.

\textbf{Training details:} We validate the effectiveness of our method with scaling factor 4 following the previous works \cite{srgan, esrgan}. The data augmentation and discriminator networks used in SRGAN and ESRGAN are adopted in our SRResNet- and RRDB-based models, respectively, for a fair comparison. The size of training input patches is set as 32×32. For both the hybrid EA-Adam optimizer and the attention-based weight regression network, the learning rate is set as $1e^{-5}$ for the SRResNet backbone, and $1e^{-4}$ for the RRDB and SwinIR backbones. To balance the performance and the training efficiency, we set the population size as $N=5$. The $T^{Adam}$ for the Adam optimizer is 10, the $T^{EA}$ for the EA optimizer is 1, and the total optimization epoch $T$ is 100. The probability $\delta$ of selecting parents from the neighboring population $\mathcal{B}$ is 0.7. 

\begin{table*}
	\renewcommand{\arraystretch}{1.2}
	\caption{
        \begin{justify}
        \justifying
Quantitative comparison between the state-of-the-art methods and the proposed hybrid EA-Adam optimization method. Three groups of methods trained with the same backbone are compared, where SRResNet \cite{srgan}, RRDB \cite{esrgan} and SwinIR \cite{swinir} backbones are used, respectively. The `SwinIR-$L_{GAN}$’ is the SwinIR model trained with the $L_{GAN}$ loss by the Adam optimizer. The best results of each group are highlighted in bold. 
        \end{justify}
 }
	\label{table3}
	\centering
	\scriptsize
        \resizebox{\linewidth}{!}{
        \begin{tabular}{c|c|ccc|cccccc|cc}
		\hline
		\multicolumn{2}{c|}{Method}& \makecell[c]{SRGAN\\{\cite{srgan}}}& \makecell[c]{RankSR-\\GAN{ \cite{ranksrgan}}}&\makecell[c]{SRResNet\\+ Ours}&\makecell[c]{ESRGAN\\\cite{esrgan}}&\makecell[c]{SPSR\\\cite{spsr}}&\makecell[c]{LDL\\\cite{LDL}}&\makecell[c]{CAL-\\GAN \cite{iccv2023}}&	\makecell[c]{WGSR\\\cite{WGSR}}&\makecell[c]{RRDB\\+ Ours}&\makecell[c]{SwinIR\\+ $L_{GAN}$}&\makecell[c]{SwinIR\\+ Ours}\\
		\hline
        \multicolumn{2}{c|}{Training Dataset}& DIV2K& DIV2K&DIV2K&DF2K+OST& DIV2K&DIV2K&DIV2K&DIV2K&DIV2K&DIV2K&DIV2K\\
		\hline
		\multirow{4}{*}{Set5}&PSNR $\uparrow$&29.93&29.77&\textbf{30.25}&30.45&30.40&30.99&31.04&30.56&\textbf{31.31}&30.35&\textbf{30.40}\\
		&SSIM$\uparrow$&0.8535&0.8394&\textbf{0.8584}&0.8582&0.8443&0.8679&0.8555&0.8529&\textbf{0.8792}&0.8543 &\textbf{0.8581}\\
		&LPIPS$\downarrow$&0.0740&0.0763&\textbf{0.0694}&0.0741&0.0686&\textbf{0.0643}&0.0670&0.0741&0.0683&0.0694&\textbf{0.0678}\\
  &DISTS$\downarrow$&0.0999&\textbf{0.0958}&0.0975&0.0972&0.0925&\textbf{0.0938}&0.1007&0.1073&0.0947&0.0960&\textbf{0.0953}\\
		\hline
		\multirow{4}{*}{Set14}&PSNR$\uparrow$&26.53&26.47&\textbf{26.98}&27.07&26.64&27.20&27.31&26.97&\textbf{27.42}&26.61&\textbf{26.78}\\
		&SSIM$\uparrow$&0.7173&0.7031&\textbf{0.7342}&0.7085&0.7138&0.7429&0.7367&0.7255&\textbf{0.7537}&0.7253&\textbf{ 0.7302}\\
		&LPIPS$\downarrow$&0.1430&0.1398&\textbf{0.1330}&0.1315&0.1345&0.1301&0.1308&0.1372&\textbf{0.1239}&0.1268&0.1268\\
  &DISTS$\downarrow$&0.1113&0.1105&\textbf{0.1043}&\textbf{0.0985}&0.0990&0.1003&0.1127&0.1147&0.1021&0.1074&\textbf{0.1047}\\
		\hline
		\multirow{4}{*}{BSD100}&PSNR$\uparrow$&25.58&25.51&\textbf{25.94}&25.33&25.51&26.12&26.29&26.01&\textbf{26.51}&25.58&\textbf{25.75}\\
		&SSIM$\uparrow$&0.6677&0.6510&\textbf{0.6836}&0.6643&0.6599&0.6941&0.6829&0.6815&\textbf{0.7071}&0.6764&\textbf{0.6832}\\
		&LPIPS$\downarrow$&0.1770&0.1829&\textbf{0.1766}&0.1602&0.1665&0.1614&0.1672&0.1791&\textbf{0.1609}&0.1568&\textbf{0.1560}\\
   &DISTS$\downarrow$&0.1293&0.1290&\textbf{0.1286}&\textbf{0.1174}&0.1186&0.1243&0.1287&0.1351&0.1212&0.1223&\textbf{0.1207}\\
		\hline
		\multirow{4}{*}{Manga109}&PSNR$\uparrow$&28.12&27.94&\textbf{28.67}&28.42&28.56&\textbf{29.41}&29.21&28.13&29.33&28.60&\textbf{28.64}\\
		&SSIM$\uparrow$&0.8656&0.8500&\textbf{0.8735}&0.8619&0.8590&0.8767&0.8671&0.8521&\textbf{0.8874}&0.8650 &\textbf{0.8694}\\
		&LPIPS$\downarrow$&0.0704&0.0755&\textbf{0.0648}&0.0644&0.0664&0.0547&0.0690&0.0760&\textbf{0.0536}&0.0631&\textbf{0.0621}\\
  &DISTS$\downarrow$&0.0551&0.0560&\textbf{0.0540}&0.0467&0.0460&0.0402&0.0498&0.0653&\textbf{0.0398}&0.0407&\textbf{0.0391}\\
		\hline 
		\multirow{4}{*}{Urban100}&PSNR$\uparrow$&24.41&24.53&\textbf{24.83}&24.36&24.80&25.50&25.33&24.83&\textbf{25.57}&25.06 &\textbf{25.14}\\
		&SSIM$\uparrow$&0.7320&0.7284&\textbf{0.7479}&0.7363&0.7473&0.7692&0.7639&0.7453&\textbf{0.7752}&0.7564&\textbf{0.7594}\\
		&LPIPS$\downarrow$&0.1439&0.1435&\textbf{0.1354}&0.1235&0.1206&\textbf{0.1096}&0.1167&0.1303&0.1150&0.1215&\textbf{0.1193}\\
  &DISTS$\downarrow$&0.1076&0.1062&\textbf{0.1038}&0.0880&0.0861&0.0861&0.0877&0.1065&\textbf{0.0842}&0.0895
&\textbf{0.0863}\\
		\hline
		\multirow{3}{*}{DIV2K100}&PSNR$\uparrow$&28.17&28.07&\textbf{28.62}&28.18&28.18&28.96&28.95&28.98&\textbf{29.34}&28.63&\textbf{ 28.77}\\
		&SSIM$\uparrow$&0.7765&0.7654&\textbf{0.7900}&0.7779&0.7720&0.7970&0.7897&0.7980&\textbf{0.8123}&0.7897&\textbf{0.7945}\\
		&LPIPS$\downarrow$&0.1254&0.1318&\textbf{0.1208}&0.1151&0.1126&\textbf{0.1008}&0.1072&0.1191&0.1038&0.1044&\textbf{0.1027}\\
  &DISTS$\downarrow$&0.0663&0.0657&\textbf{0.0646}&0.0594&0.0546&\textbf{0.0529}&0.0600&0.0687&0.0554&0.0562
&\textbf{0.0537}\\
		\hline
         \multicolumn{2}{c|}{\#FLOPs}&\multicolumn{3}{c|}{10.38G}&\multicolumn{6}{c|}{73.43G}&\multicolumn{2}{c}{4.42G}\\
        \hline
        \multicolumn{2}{c|}{\#Params}&\multicolumn{3}{c|}{1.52M}&\multicolumn{6}{c|}{16.69M}&\multicolumn{2}{c}{0.93M}\\
        \hline
	\end{tabular}}
\end{table*}

\begin{table*}
	\renewcommand{\arraystretch}{1.2}
	\caption{
        \begin{justify}
        \justifying
Quantitative comparisons between the state-of-the-art perceptually realistic SR models and the models trained by our method. The 'RRDB+Ours-PD' employs the RRDB backbone and the weight setting in PD-ADMM \cite{zhang2022perception}. The 'DualFormer+Ours' employs the RRDB backbone and the discriminator in DualFormer \cite{DualFormer}. The best results of each group are highlighted in bold.
        \end{justify}}
	\label{table31}
	\centering
        \scriptsize
        \resizebox{\linewidth}{!}{
        \begin{tabular}{c|c|ccc|cc|cccc}
	    \hline
		\multicolumn{2}{c|}{Method}&\makecell[c]{G-MGBP\\{~\cite{2018ECCVW}}}&\makecell[c]{CFSNet\\\cite{cfsnet}}&\makecell[c]{SRResNet\\+Ours}&\makecell[c]{PD-AD\\MM\cite{zhang2022perception}}&\makecell[c]{RRDB\\+Ours-PD}&\makecell[c]{SROOE\\~\cite{SROOE}}&\makecell[c]{DualFormer\\~\cite{DualFormer}}&\makecell[c]{RRDB\\+Ours}&\makecell[c]{DualFormer\\+Ours}\\ \hline
        \multicolumn{2}{c|}{Training Dataset}&DIV2K&DIV2K&DIV2K&DIV2K&DIV2K&DF2K&DF2K&DF2K&DF2K\\
		\hline
		\multirow{4}{*}{Set5}&PSNR$\uparrow$&29.54&30.23&\textbf{30.25}&\textbf{31.82}&31.68&31.30&31.35&\textbf{31.45}&31.32\\
        &SSIM$\uparrow$&0.8419&0.8503&\textbf{0.8584}&0.8810&\textbf{0.8838}&0.8671&0.8690&\textbf{0.8744}&0.8713\\
		&LPIPS$\downarrow$&0.0855&0.0781&\textbf{0.0694}&0.0758&\textbf{0.0746}&0.0646&0.0678&0.0640&\textbf{0.0595}\\
   &DISTS$\downarrow$&0.1167&0.0986&\textbf{0.0975}&0.1107&\textbf{0.1047}&0.0936&0.0920&0.0930&\textbf{0.0917}\\
		\hline
		\multirow{4}{*}{Set14}&PSNR$\uparrow$&26.58&26.59&\textbf{26.98}&\textbf{27.99}&27.95&27.36&27.47&\textbf{27.59}&27.53\\
		&SSIM$\uparrow$&0.7151&0.7170&\textbf{0.7342}&0.7589&\textbf{0.7668}&0.7362&0.7397&\textbf{0.7479}&0.7443\\
		&LPIPS$\downarrow$&0.1502&0.1619&\textbf{0.1330}&0.1401&\textbf{0.1336}&0.1167&0.1191&0.1137&\textbf{0.1134}\\
  &DISTS$\downarrow$&0.1207&0.1178&\textbf{0.1043}&0.1129&\textbf{0.1083}&\textbf{0.0907}&0.0933&0.1007&0.0974\\
		\hline
		\multirow{4}{*}{BSD100}&PSNR$\uparrow$&25.68&25.60&\textbf{25.94}&26.86&\textbf{26.93}&26.34&26.50&\textbf{26.64}&26.43\\
		&SSIM$\uparrow$&0.6673&0.6677&\textbf{0.6836}&0.7137&\textbf{0.7206}&0.6953&0.6895&\textbf{0.7044}&0.7019\\
		&LPIPS$\downarrow$&0.1882&0.1822&\textbf{0.1766}&0.1854&\textbf{0.1774}&0.1528&0.1583&0.1583&\textbf{0.1520}\\
  &DISTS$\downarrow$&0.1418&0.1312&\textbf{0.1286}&0.1369&\textbf{0.1312}&\textbf{0.1172}&0.1207&0.1209&0.1176\\
		\hline
		\multirow{4}{*}{Manga109}&PSNR$\uparrow$&28.17&28.54&\textbf{28.67}&\textbf{30.17}&29.92&\textbf{29.97}&29.82&29.82&29.87\\
	&SSIM$\uparrow$&0.8609&0.8622&\textbf{0.8735}&0.8864&\textbf{0.8952}&0.8802&0.8837&\textbf{0.8943}&0.8876\\
		&LPIPS$\downarrow$&0.0786&0.0689&\textbf{0.0648}&0.0616&\textbf{0.0528}&0.0504&0.0528&\textbf{0.0491}&0.0504\\
  &DISTS$\downarrow$&0.0726&0.0558&\textbf{0.0550}&0.0544&\textbf{0.0455}&0.0402&0.0377&0.0386&\textbf{0.0365}\\
		\hline 
		\multirow{4}{*}{Urban100}&PSNR$\uparrow$&24.40&24.72&\textbf{24.83}&\textbf{26.29}&25.89&25.94&25.68&25.94&\textbf{25.95}\\
		&SSIM$\uparrow$&0.7415&0.7445&\textbf{0.7479}&\textbf{0.7882}&0.7824&0.7801&0.7741&0.7853&\textbf{0.7854}\\
		&LPIPS$\downarrow$&0.1477&0.1362&\textbf{0.1354}&0.1234&\textbf{0.1220}&\textbf{0.1080}&0.1150&0.1124&0.1089\\
  &DISTS$\downarrow$&0.1358&0.1086&\textbf{0.1038}&0.1016&\textbf{0.0931}&0.0846&0.0842&0.0833&\textbf{0.0827}\\
		\hline
		\multirow{4}{*}{DIV2K100}&PSNR$\uparrow$&28.27&28.49&\textbf{28.62}&29.71&\textbf{29.91}&29.24&29.30&29.33&\textbf{29.36}\\
		&SSIM$\uparrow$&0.7766&0.7826&\textbf{0.7900}&0.8117&\textbf{0.8226}&0.8016&0.8020&\textbf{0.8142}& 0.8111\\
		&LPIPS$\downarrow$&0.1488&0.1233&\textbf{0.1208}&0.1223&\textbf{0.1136}&\textbf{0.0968}&0.1027&0.1034&0.0989\\
  &DISTS$\downarrow$&0.0925&0.0649&\textbf{0.0666}&0.0763&\textbf{0.0654}&0.0565&0.0551&0.0540&\textbf{0.0524}\\
		\hline
        \multicolumn{2}{c|}{\#FLOPs}&22.42G&19.22G&10.38G&109.78G&73.43G&90.06G&73.43G&73.43G&73.43G\\
        \hline
        \multicolumn{2}{c|}{\#Params}&0.28M&5.00M&1.52M&18.67M&16.69M&103.44M&16.69M&16.69M&16.69M\\
        \hline
	\end{tabular}}
 \vspace{-3pt}
\end{table*}

\begin{figure*}
	\centering 
	\includegraphics[scale=0.68]{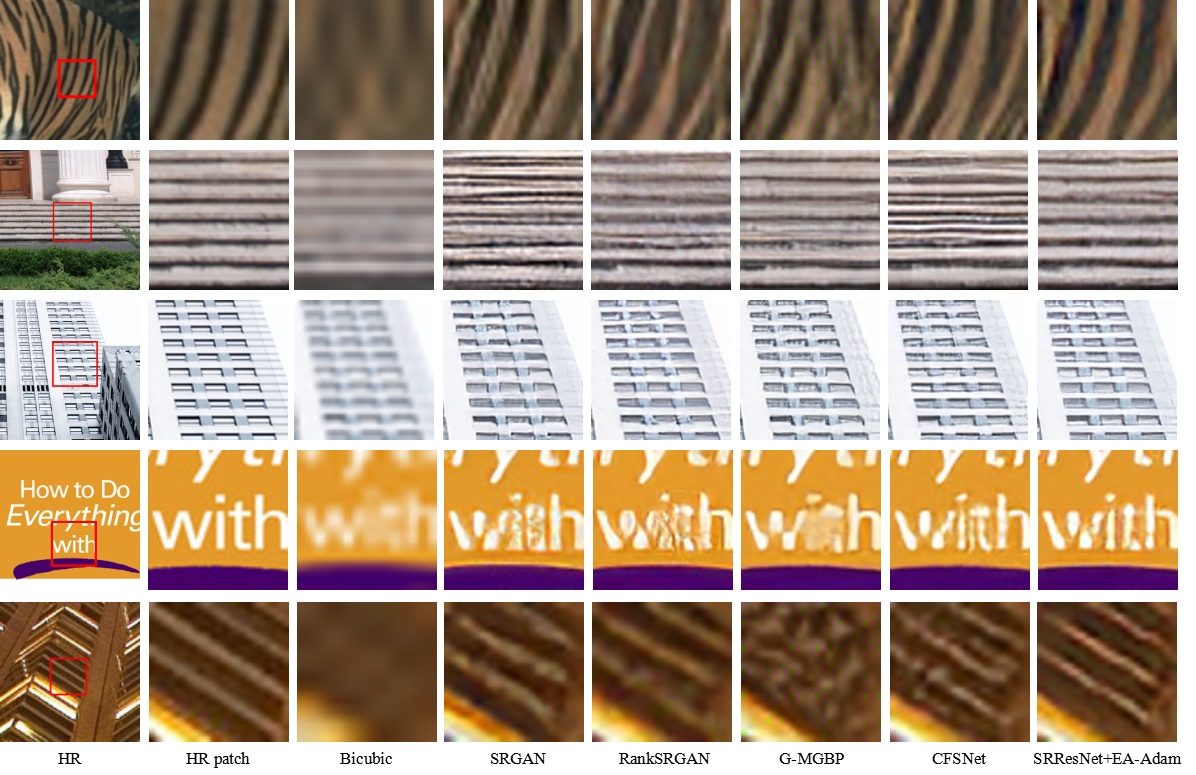}
	\caption{Visual comparisons among the proposed EA-Adam optimization method and other state-of-the-art methods under the SRResNet backbone. }	
	\label{compare_srresnet}
\end{figure*}
\begin{figure*}
	\centering 
	\includegraphics[scale=0.67]{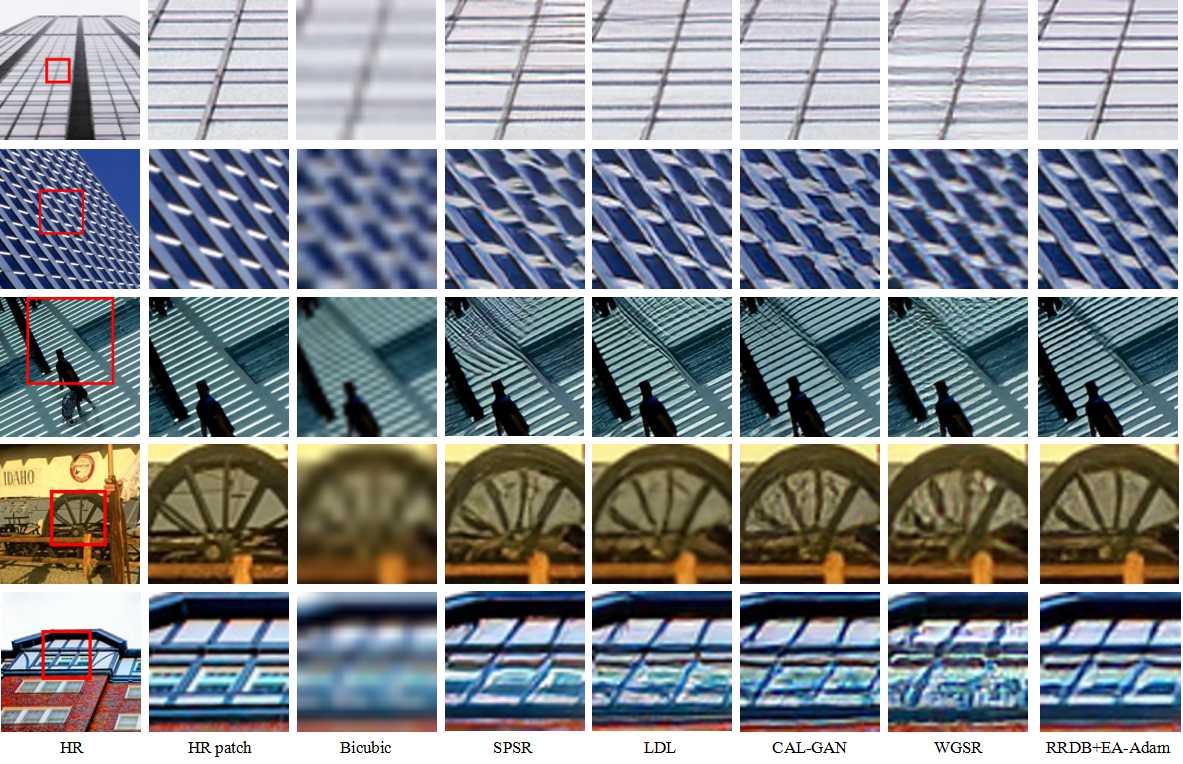}
	\caption{Visual comparisons among the proposed EA-Adam optimization method and other state-of-the-art methods under the RRDB backbone.}	
	\label{compare_rrdb}
\end{figure*}

\begin{figure*}
	\centering 
	\includegraphics[scale=0.74]{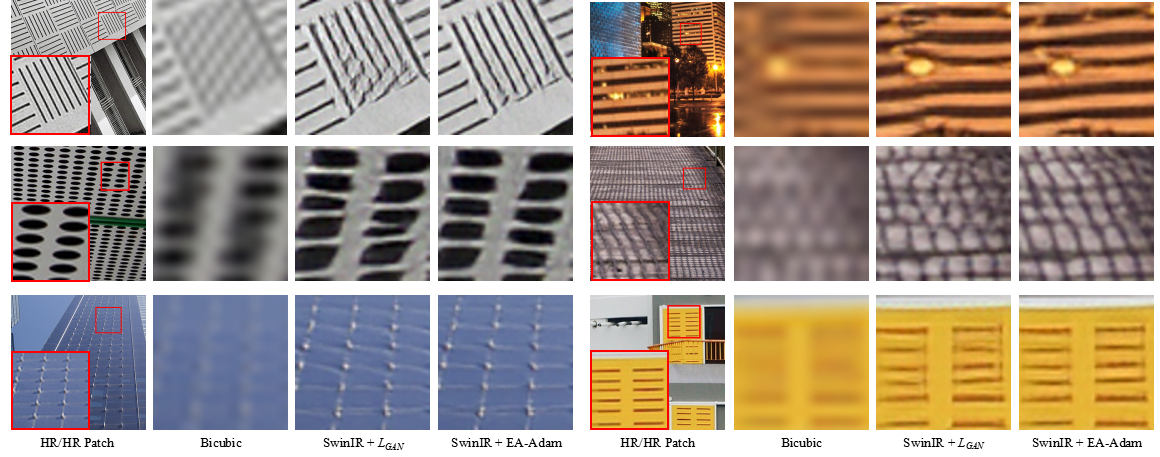}
	\caption{Visual comparisons among the SwinIR optimized with the $L_{GAN}$ loss by Adam optimizer and the proposed EA-Adam optimization method.}	
	\label{compare_swinir}
\end{figure*}

\begin{figure*}
	\centering 
	\includegraphics[scale=0.74]{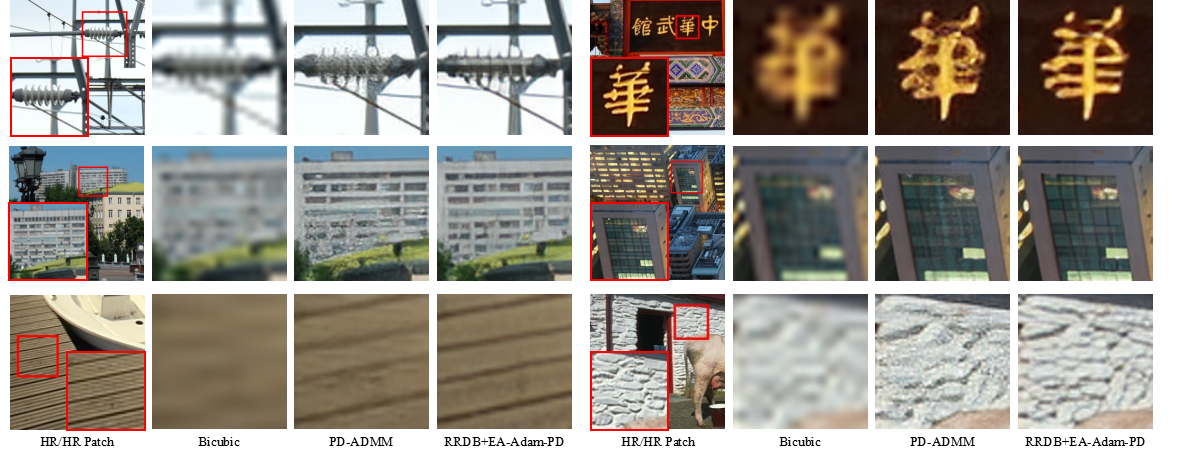}
	\caption{Visual comparisons among the proposed EA-Adam optimization method and PD-ADMM.}	
	\label{compare_pdadmm}
\end{figure*}

\begin{figure*}
	\centering 
	\includegraphics[scale=0.9]{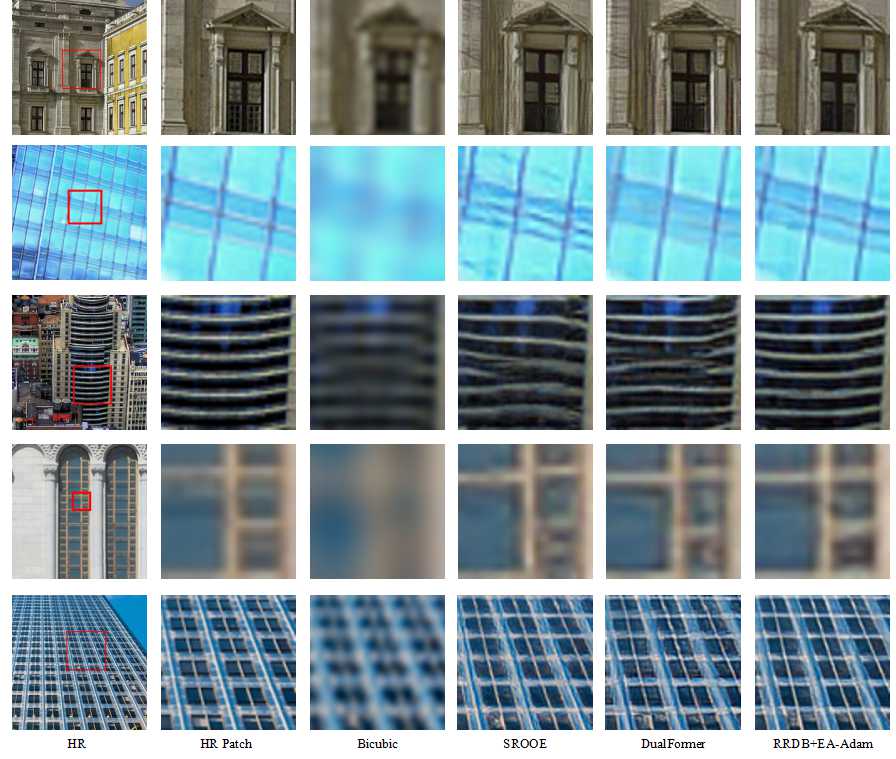}
	\caption{Visual comparisons among the proposed EA-Adam optimization method with SROOE and DualFormer.}	
	\label{compare_new}
\end{figure*}

\subsection{Comparison with State-of-the-Arts}

In this section, we compare the final fused model by our proposed hybrid EA-Adam optimizer and network fusion method against state-of-the-art methods. Table \ref{table3} provides the quantitative comparisons among the methods with the same backbone. Table \ref{table31} provides the quantitative comparisons among the methods with similar model capacity.

From Table \ref{table3}, we can see that for the lightweight network models with the SRResNet backbone, SRGAN gets better PSNR and SSIM scores than RankSRGAN on almost all the benchmarks, while they have higher or lower LPIPS scores on different benchmarks. Our proposed method improves the fidelity- and perceptual-oriented metrics by a large margin under all benchmarks without increasing the SR model complexity. This validates the advantages of our hybrid EA-Adam optimization strategy, which could push the PF toward the optimal point of solution space. For models trained with the RRDB backbone, we can see that our proposed method achieves better results than the competing methods ESRGAN, SPSR, LDL, CAL-GAN, and WGSR in most cases. The LDL also demonstrates competing results in several instances, especially for the perceptual quality metric on the Urban100 dataset. This advantage comes from its iterative and local discriminative strategy that can effectively inhibit visual artifacts and encourage the generation of high-frequency details. Furthermore, our method with the SwinIR backbone demonstrates significant improvement over its counterparts, which verifies that the proposed method can be applied to both CNN-based and transformer-based models.

From Table \ref{table31}, we can see that the proposed method achieves consistent improvement on most benchmarks in terms of both perceptual quality (LPIPS) and reconstruction accuracy (PSNR and SSIM), although the compared methods such as G-MGBP, CFSNet and SROOE have more model parameters and FLOPs (\eg, 22.42G FLOPs and 19.22G FLOPs for `G-MGBP' and `CFSNet', while 10.38G FLOPs for `SRResNet+Ours'). 
In addition, we test the released model of PD-ADMM, which performs favorably in terms of distortion control yet sacrifices the perceptual quality. 
PD-ADMM applies the same weight, \ie, [0.5, 0.5], for perception and distortion objectives. A constraint term is further introduced to ensure that the low-frequency information in two stages is equal. The ADMM optimization is applied to solve the constrained optimization problem in PD-ADMM. Since it is improper and difficult to integrate the constraint term into our proposed hybrid EA-Adam approach with $N$-optimized models, to fairly compare our method with PD-ADMM, we apply the optimization objective of PD-ADMM to EA-Adam without its constraint. One can see that our method outperforms PD-ADMM in almost all cases but with fewer FLOPs and parameters.
SROOE employs an additional condition network to learn the loss aggregation weights, leading to significantly larger model size and Flops than the other methods.
In contrast, our EA-Adam method is more efficient while achieving comparable results with SROOE.
Our method obtains better performance than DualFormer on both perception and distortion metrics. Note that DualFormer employs a spectral and a spatial discriminator. Our EA-Adam presents a new training strategy from the optimization perspective, which can be applied to DualFormer to further enhance its performance, as shown in `DualFormer+Ours’ in Table \ref{table31}.

We provide 4× SR visual comparisons between models (with SRResNet \cite{srgan}, RRDB \cite{esrgan} and SwinIR \cite{swinir} backbones) trained by our method and other state-of-the-arts models in Fig. \ref{compare_srresnet}, Fig. \ref{compare_rrdb} and Fig. \ref{compare_swinir}. We do not compare with the early published method ESRGAN due to limited space. In addition, we give the visual comparisons between PD-ADMM \cite{zhang2022perception} and `RRDB+Ours-PD', \ie, the RRDB model trained by our EA-Adam optimization method with the weight setting in PD-ADMM, in Fig. \ref{compare_pdadmm}. 
The visual comparisons between SROOE \cite{SROOE}, DualFormer \cite{DualFormer} and our proposed EA-Adam optimization method are presented in Fig. \ref{compare_new}.
One can see that our method can restore more correct visual patterns (\eg, the steps in the second group of Fig. \ref{compare_srresnet} and the zebra crossings in the third group of Fig. \ref{compare_rrdb}) while inhibiting the artifacts (\eg, the building windows in the fifth group of Fig. \ref{compare_new}). 
At the same time, our model can generate many high-frequency details (\eg, the lines of buildings and wood floor in the first and third group of Figs. \ref{compare_swinir} and \ref{compare_pdadmm}), respectively.


\subsection{Ablation Studies}
In this section, we conduct a series of ablation studies to validate the effectiveness of the proposed hybrid EA-Adam optimization. We first conduct experiments to compare our hybrid EA-Adam algorithm with the Adam algorithm, and compare their results with different selections of $N$. Then we compare our network fusion method with the learnable weight method. Furthermore, we study the selection of different loss functions in optimizing the weight regression network. We also study the aggregation weights of obtained models with different perception-distortion preferences. All ablation results are evaluated on the Urban100 dataset.

\begin{table}
	\renewcommand{\arraystretch}{1.2}
	\caption{
        \begin{justify}
        \justifying
 Quantitative comparison between the models optimized by Adam, AdamW, our hybrid EA-Adam, and hybrid EA-AdamW optimizers. The SRResNet backbone is adopted and the models are evaluated on the Urban100 dataset. `Fused Model' is obtained by applying our network fusion method to the four trained models. The best results are highlighted.
        \end{justify}}
    \vspace{3pt}
	\label{table2}
	\centering
        \scriptsize
	\begin{tabular}{p{0.5cm}|p{1.3cm}<{\centering}|p{0.9cm}<{\centering}p{0.9cm}<{\centering}p{0.9cm}<{\centering}p{0.6cm}<{\centering}p{0.5cm}<{\centering}}
		\hline
		\multicolumn{2}{c|}{Model}&Model1&Model2&Model3&Model4&Fused model\\
		\hline
		\multicolumn{2}{c|}{Weights}&$[0.25,0.75]$&$[0.5,0.5]$&$[0.75,0.25]$&$[1,0]$&-\\
		\hline
		\multirow{4}{*}{PSNR}&Adam&24.36&23.82&23.96&24.22&24.41\\
        &EA-Adam&\textbf{25.78}&\textbf{25.75}&\textbf{24.95}&\textbf{24.73}&\textbf{24.83}\\
        \cline{2-7}
        &AdamW&25.48&24.67&23.41&24.18&24.37\\	
        &EA-AdamW&\textbf{25.95}&\textbf{25.81}&\textbf{25.21}&\textbf{24.82}&\textbf{24.91}\\
		\hline
		\multirow{4}{*}{LPIPS}&Adam&0.3061&0.2899&0.2665&0.1446&0.1437\\
        &EA-Adam&\textbf{0.2120}&\textbf{0.2055}&\textbf{0.1536}&\textbf{0.1384}&\textbf{0.1354}\\
        \cline{2-7}
	&AdamW&0.2570&0.2925&0.2306&0.1477&0.1444\\	
        &EA-AdamW&\textbf{0.2265}&\textbf{0.2201}&\textbf{0.1519}&\textbf{0.1404}&\textbf{0.1368}\\
		\hline
	\end{tabular}
\vspace{-3pt}
\end{table}

\begin{figure*}
	\centering 
	\includegraphics[scale=1.1]{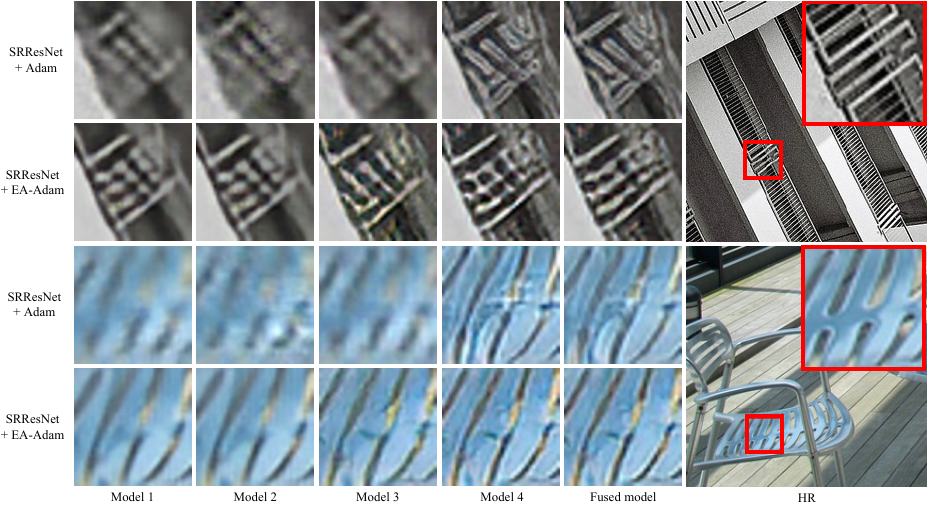}
	\caption{Visual comparisons among SR models trained by the original Adam optimizer and our hybrid EA-Adam optimizer with the SRResNet \cite{srgan} backbone. Each model trained by EA-Adam outperforms the one trained by Adam in both fidelity and perceptual quality. The fused model by the proposed EA-Adam method achieves further benefits in fidelity and perception.}	
	\label{fig5}
 \vspace{+3mm}
\end{figure*}

\noindent\textbf{Hybrid EA-Adam vs. Adam.}
\label{EAvsADAM}
We compare the performance of models optimized by the original Adam and our hybrid EA-Adam. In addition, we replace Adam with AdamW \cite{Adamw}, a recently developed variant of Adam, and report the results of AdamW and hybrid EA-AdamW for a more comprehensive evaluation of our proposed method. 
As traditional optimizers (Adam and AdamW) cannot directly optimize a multi-objective problem, we uniformly set 4 combinations of loss weights, \ie, $[0.25,0.75]$, $[0.5,0.5]$, $[0.75,0.25]$ and $[1,0]$, to weigh the two objectives in Eq.~\eqref{Eq2} into a single-objective optimization problem. For example, if the weight is $[0.25,0.75]$, the optimization objective of Adam is $0.25 \times f_1 + 0.75 \times f_2$, which is the same as that of the Adam steps in our proposed hybrid optimizer. The models optimized by different weights are named `Model1' to `Model4', respectively. In addition, we apply the same network fusion method (as in Section 3.3) to the 4 models optimized by Adam. 

Table \ref{table2} gives the quantitative comparison. Without loss of generality, we conduct experiments with the SRResNet backbone and evaluate on the Urban100 dataset. PSNR and LPIPS metrics are used to measure the fidelity and perceptual quality, respectively. 
One can see that our hybrid EA-Adam/AdamW optimizers outperform Adam/AdamW by a large margin in most cases. Both the fidelity and perceptual quality metrics are improved, validating that the proposed hybrid EA-Adam strategy can push the PF closer to the optimal point of the metric space. For most models, EA-Adam improves PSNR by more than 0.4dB over Adam and improves LPIPS by more than $10\%$. This might be caused by the gradient vanishing problem in the gradient-based optimization process so that the models can be trapped into a local minimum. In contrast, with our hybrid EA-Adam optimizer, effective interaction between different models can be introduced by the crossover operator in EA, so that the models can be stably optimized with better convergence.

On the other hand, each of the four models may perform well in one aspect, yet sacrifices the other aspect as a trade-off. The fused model by our network fusion method achieves a better balance between fidelity and perceptual quality. Specifically, although the fidelity indices of the fused model are lower than that of Models 1, 2, and 3, its perceptual performance is better than Model 4, which is trained to maximize perceptual quality. This result well aligns with the goal of fusion network, which aims to further improve the perceptual quality of the fused model without reducing much the fidelity.
It is worth noting that the bad convergence of Models 1 to 3 makes the traditional Adam/Adamw optimizer gain less improvement from the network fusion process. In contrast, the proposed hybrid optimization strategy facilitates the effectiveness of network fusion via the interaction of individuals in crossover operation. In addition, we find that AdamW does not show many advantages over Adam in the SR task. This is because AdamW is mainly proposed to handle the over-fitting problem in high-level vision tasks by introducing weight decay. However, this may not fit the SR task as the optimization might be under-fitting to the large space of image details.  

Fig. \ref{fig5} shows the visual comparisons among SR models trained by Adam and our EA-Adam optimizers, where consistent observation with Table~\ref{table2} can be drawn. Specifically, models optimized by EA-Adam outperform their counterparts optimized by Adam, where the structures and patterns are more loyal to the HR and richer details are generated. It can also be observed that while each of the four models is preferred in either fidelity or perceptual quality, the fused model by our method shows comprehensive enhancement in both aspects.

\begin{table}[t]\scriptsize
\renewcommand{\arraystretch}{1.2}
\caption{
        \begin{justify}
        \justifying 
        Training time and quantitative comparisons on Urban100 benchmark by using different numbers of individuals ($N$) in EA.
        \end{justify}}
\vspace{-0.5em}
\label{Ablation_N}
\centering
\begin{tabular}{c|ccc|c}
\hline
N & PSNR  & SSIM   & LPIPS & Training time \\ \hline
3 & 25.34 & 0.7525 & 0.1184 & 63h\\
5 & 25.57 & 0.7752 & 0.1150&  112h\\
7 & 25.60 & 0.7804 & 0.1125&  160h\\
9 & 25.64 & 0.7823 & 0.1114&  211h\\ \hline
\end{tabular}
\end{table}

\begin{table}
	\renewcommand{\arraystretch}{1.2}
        \vspace{-1.5em}

        \caption{
        \begin{justify}
        \justifying
        Quantitative comparison between the `Learnable Weight', `Weight Regression', and `Our Fusion Method' on the Urban100 dataset. The RRDB backbone is adopted. The best and second results are highlighted in \textcolor{red}{red} and \textcolor{blue}{blue}.
        \end{justify}
        }
        
	\label{table: weight}
        \scriptsize
	\centering
	\begin{tabular}{p{2.7cm}m{1.5cm}m{1.5cm}m{1.5cm}m{1.5cm}}
             \hline
            &PSNR&SSIM&LPIPS\\
            \hline
		Learnable Weight&\textcolor{blue}{25.60}&0.7748&0.1195\\
            Weight Regression &\textcolor{red}{25.71}&\textcolor{red}{0.7756}&\textcolor{red}{0.1124}\\
            Our Fusion Method &25.57&\textcolor{blue}{0.7752}&\textcolor{blue}{0.1150}\\
            \hline
	\end{tabular}
\end{table}

\noindent\textbf{Selection of the number of optimized models $N$.} To show the influence of the number of optimized models, we conduct experiments by using different selections of $N$. The results on Urban100 are shown in Table \ref{Ablation_N}. One can see that increasing $N$ can obtain better perception-distortion performance but at the price of longer optimization time. When
$N$ is greater than 5, the benefits are not significant. Considering both performance and training time, we choose $N = 5$ in the experiments.

\noindent\textbf{Network Fusion vs. Learnable Weight.}
We propose a two-stage framework for fusing $N$ networks. The first stage is the attention-based weight regression, which generates a set of weights for each input LR image. Therefore, the corresponding weight vectors of validation data can be obtained. The second stage is model fusion, which obtains a single stronger model by averaging all the weight vectors of validation data. To show the effectiveness of the proposed fusion method, we introduce a fusion method by setting the fusion weights of $N$ models as learnable parameters directly, which is named `Learnable Weight'. We compare our two stages, named `Weight Regression' and `Our Fusion Method', with `Learnable Weight' on the Urban100 dataset in Table \ref{table: weight}. We can see that the `Weight Regression' achieves the best results in all metrics since it can dynamically infer the optimal weights for each input based on image content. `Our Fusion Method' obtains better performance than `Learnable Weight' in perceptual quality metric LPIPS with 0.0045 improvement while keeping comparable PSNR and SSIM values. This is because the universal network weight is inherited from `Weight Regression' using validation data, which holds a stronger generalization capability.

\begin{table}
	\renewcommand{\arraystretch}{1.2}
        \caption{
        \begin{justify}
        \justifying
        Quantitative comparison on Urban100 benckmark among different loss functions for the network fusion process. The RRDB backbone is adopted.
        \end{justify}}
	\label{table: loss}
	\centering
        \scriptsize
	\begin{tabular}{p{2.7cm}m{1.8cm}m{1.8cm}m{0.7cm}}
             \hline
            &PSNR&SSIM&LPIPS\\
            \hline
            RRDB+Ours-$\ell_1$&26.55&0.8029&0.1955\\
		RRDB
            +Ours-PD [65]&25.89&0.7824&0.1220\\
            RRDB
            +Ours-$L_{\scalebox{.85}{adv}}$&25.57&0.7752&0.1150\\
            \hline
	\end{tabular}
\end{table}

\noindent\textbf{Loss Selection for Network Fusion.}
The fusion network aims to leverage the strength of $N$ models for perceptually more plausible SR results. Table \ref{table: loss} gives the numerical comparisons on the Urban100 benchmark with $\ell_1$, $PD$ \cite{zhang2022perception} and $L_{adv}$ loss functions. We choose the adversarial loss in this paper because it achieves the best LPIPS score, which implies perceptually more realistic outputs.

\begin{table}
\renewcommand{\arraystretch}{1.2}
\setlength{\tabcolsep}{1pt}
	\caption{
        \begin{justify}
        \justifying
         Aggregation weights of two layers (L1 and L2) in the RRDB models with different perception-distortion preferences given two inputs (I1 and I2), and the final fusion model (Ours).
        \end{justify}}
	\label{table: layers}
	\centering
        \scriptsize
	\begin{tabular}{p{0.4cm}p{4.2cm}<{\centering}p{4.2cm}<{\centering}}
             \hline
            &L1&L2\\
            \hline
            I1&[0.0463, 0.0167, 0.0678, 0.3333, 0.5358]&[0.2348, 0.2084, 0.1426, 0.1452, 0.2689]\\
            I2&[0.0263, 0.0144, 0.1014, 0.3307, 0.5272]&[0.3371, 0.1061, 0.0022, 0.1985, 0.3561]\\
            Ours&[0.0965, 0.0733, 0.1429, 0.2629, 0.4243]&[0.3203, 0.1212, 0.0224, 0.1932, 0.3428]\\
            \hline
	\end{tabular}
\end{table}

\begin{table}[t]
\renewcommand{\arraystretch}{1.2}
	\caption{
        \begin{justify}
        \justifying
        Quantitative comparisons on Urban100 benchmark between different ways of aggregation weight combination. The RRDB backbone is adopted. The best results are highlighted in bold.
        \end{justify}}
	\label{table: weights}
	\centering
        \scriptsize
	\begin{tabular}{p{2.3cm}m{1.8cm}m{1.8cm}m{0.7cm}}
             \hline
            &PSNR&SSIM&LPIPS\\
            \hline
		(a)&21.86&0.6546&0.2269\\
            (b)&24.52&0.7573&0.1755\\
            (c)&\textbf{25.57}&\textbf{0.7752}&\textbf{0.1150}\\
            \hline
	\end{tabular}
\end{table}

\noindent\textbf{Aggregation Weights of Obtained Models.}
In the fusion stage, the network parameters of the $N$ models are frozen, and the aggregation weights are optimized for each layer using a set of validation data to fuse the $N$ models into one by interpolation, as done in those network interpolation methods \cite{network-interpolation}.
To further show the advantage of our proposed fusion method, we study the aggregation weights of obtained models. Given two inputs (I1 and I2), the aggregation weights in two random layers (L1 and L2) of the $5$ obtained RRDB models are shown in Table \ref{table: layers}. 
The aggregation weights of the final fusion model `Ours' are also given. 
It can be seen that the weights are neither evenly distributed among models nor concentrated on one or two models. 
The aggregation weight distribution varies in layers, and each model contributes to the final fusion. 
This demonstrates the effectiveness of our EA-Adam optimizer, which can simultaneously obtain models with different focuses. 
We also perform the ablation study on aggregation weight by (a) directly averaging the parameters of the $5$ networks, (b) applying the aggregation weights of one layer to all layers of the $5$ networks, and (c) our proposed fusion strategy. The results on Urban100 are shown in Table \ref{table: weights}. 
Our fusion method achieves the best result, showing the effectiveness of our network fusion method. 

\subsection{Applications to real-world Super-Resolution}
To demonstrate the generalization capability of the proposed method, we further present the comparisons by using the complex
real-world SR setting. Specifically, we use the generator and discriminator of RealESRGAN \cite{realesrgan} and
optimize the model with our proposed EA-Adam hybrid optimizer and the fusion strategy. The DF2K \cite{div2k, flick2k} is used as the training dataset, and the LR images are degraded with the RealESRGAN degradation pipeline \cite{realesrgan}. The test data are from the real-world datasets, including RealSR \cite{realsr} and DRealSR \cite{drealsr}. The compared methods include RealESRGAN \cite{realesrgan}, IKC \cite{IKC}, LDL \cite{LDL}, IKRNet \cite{IKR-Net} and DASR \cite{DASR}. The results are shown in Table \ref{table: real}. One can see that IKC obtains the worst results since it focuses on solving blur degradation and cannot adapt to real-world SR tasks. The proposed method obtains the
best perception-distortion results in the real-world SR task, demonstrating its effectiveness.
\begin{table}[t]\scriptsize
\renewcommand{\arraystretch}{1.2}

\caption{
\begin{justify}
        \justifying
Quantitative comparisons with GAN-based state-of-the-art real-world super-resolution methods on RealSR and DrealSR test datasets. The RRDB backbone is adopted. The best result is highlighted in bold.
\end{justify}}
\label{table: real}
\centering
\begin{tabular}{c|c|c}
\hline
\multirow{2}{*}{} & RealSR              & DrealSR             \\ \cline{2-3} 
                  & PSNR/SSIM/LPIPS     & PSNR/SSIM/LPIPS     \\ \hline
RealESRGAN       & 25.69/0.7616/0.2727 & 28.64/0.8053/0.2847 \\
IKC               &            19.51/0.5088/0.4337          &      27.07/0.7429/0.3994                \\
IKR-Net           &        25.15/0.7308/0.3912             &    28.94/0.8131/0.4274                 \\\hline
LDL               & 25.28/0.7567/0.2766 & 28.21/0.8126/0.2815 \\
DASR              & 27.02/0.7708/0.3151 & 29.77/0.8263/0.3126 \\\hline
RRDB+Ours           & \textbf{27.59}/\textbf{0.7806}/\textbf{0.2621} & \textbf{30.32}/\textbf{0.8358}/\textbf{0.2802}  \\ \hline
\end{tabular}
\end{table}
\subsection{Training and Inference Time}
\noindent\textbf{Training time.} We first compare the training time of our hybrid EA-Adam optimization method (including the EA-Adam optimizer and the network fusion stage) and the original Adam optimizer by training an SR model with RRDB-backbone for ×4 super-resolution. Table \ref{table4} provides the quantitative comparison of the training time. Specifically, the Adam optimizer costs 250 epochs and 48h to converge.
Though hybrid EA-Adam costs a longer time in each epoch, it converges in much fewer epochs thanks to the effective interaction in the EA step. The hybrid EA-Adam optimization method totally costs 144h to finish the model training, which is about three times the training time of Adam. 
Most of the training time of hybrid EA-Adam is spent on the EA part. It is noted that there have been various parallel computation methods \cite{shi2022improving, shi2018ppls} to speed up the EA process based on CUDA technology, showing great potential to accelerate our method in training.

\noindent\textbf{Inference time.}
While spending more time in training, the proposed method does not introduce any extra computational and memory burden during the inference stage. The trained SR model has the same inference time as other SR models with the same backbone. 
\begin{table}[t]
	\renewcommand{\arraystretch}{1.2}
	\caption{
        \begin{justify}
        \justifying
 Comparison of the training time of our hybrid EA-Adam optimization method and the Adam method.
        \end{justify}}

	\label{table4}
	\centering
        \scriptsize
	\begin{tabular}{p{1.6cm}|p{1.0cm}<{\centering}|p{1.0cm}<{\centering}|p{1.0cm}<{\centering}|p{1.6cm}<{\centering}}
		\hline
        \multirow{2}{*}{Method}&\multirow{2}{*}{Adam}&\multicolumn{2}{c|}{EA-Adam optimizer}&\multirow{2}{*}{Network Fusion}\\
        \cline{3-4}
        &&EA-iter&Adam-iter&\\
        \hline
        Run-time/epoch&0.19h&4.76h&0.80h&0.63h\\
        \hline
        Total-epoch&250&8&92&50\\
        \hline
        Total-run-time&48h&\multicolumn{2}{c|}{112h}&32h\\
        \hline
	\end{tabular}

\end{table}

\begin{figure}[t]
	\centering 
	\includegraphics[scale=0.26]{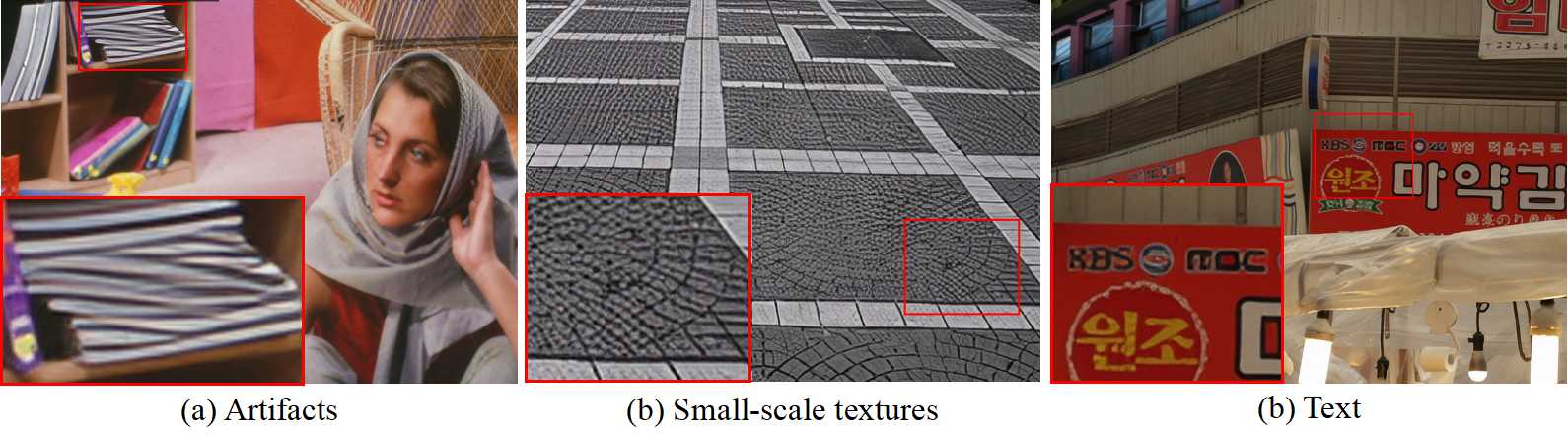}
	\caption{Visualization of failure cases generated by our model.}
	\label{failure_case}
\end{figure}

\subsection{Failure Cases}
While our method can improve the training process of GAN-based SR models and improve the model performance in perception-distortion balance, it can still fail in some challenging cases. Fig. \ref{failure_case} shows some typical examples. If the high-frequency structures are largely destroyed in the input LR image, it is hard for our model to correct them in the SR output, as shown in Fig. \ref{failure_case}(a). In addition, for some small-scale textures and text characters, our method may generate some visual artifacts, as shown in Figs. \ref{failure_case}(b)) and \ref{failure_case}(c)).

\section{Conclusion}
In this paper, we proposed a hybrid EA-Adam optimization method to optimize the perception-distortion balanced image super-resolution (SR) models. We formulated the perception-distortion trade-off as a multi-objective optimization problem. A population of SR models were optimized by alternatively performing the Adam and EA steps, where the Adam was dedicated to the model convergence and the EA focused on rescuing the model from local minimum due to its strong capability in multi-objective searching. An effective model fusion strategy was then designed to merge the trained models into a stronger one, which further improved the perception-distortion trade-off without increasing the inference cost. Extensive experiments demonstrated the effectiveness of the proposed method against the traditional Adam optimizer, and the state-of-the-art perception-distortion balanced performance of the trained SR models.

\ifCLASSOPTIONcaptionsoff
  \newpage
\fi

\footnotesize
\bibliographystyle{IEEEtran}
\bibliography{IEEEabrv,refer}

\end{document}